# Privacy and National Security Issues in Social Networks: The Challenges

Shafi'i M. Abdulhamid[1], Sulaiman Ahmad[2], Victor O. Waziri [3] and Fatima N. Jibril[4]

Department of Cyber Security Science, Federal University of Technology Minna, Nigeria.[1,2,3]
Department of Library Information Technology, Federal University of Technology Minna, Nigeria.[4]
E-mail: shafii.abdulhamid@futminna.edu.ng

*Abstract*

*Online social networks are becoming a major growth point of the internet, as individuals, companies and governments constantly desire to interact with one another, the ability of the internet to deliver this networking capabilities grows stronger. In this paper, we looked at the structure and components of the member profile and the challenges of privacy issues faced by individuals and governments that participate in social networking. We also looked at how it can be used to distort national security, how it became the new weapons of mass mobilization and also how social networks have became the rallying forces for revolutions and social justice.*

## 1. Introduction

Social network sites (SNS) are web-based services that allow people to build a public or semi-public profile within a bounded system, articulate a list of other users usually called Friends with whom they share a connection, and view and traverse their list of connections and those made by others within the system. The nature and nomenclature of these connections may vary from site to site. Social networking by definition focuses on building and reflecting personal and social relations among people who share common interests, causes or goals. A social networking site is an on-line service that attracts a community of users and provides such users with a variety of tools for posting personal data and creating user-generated content directed to a given user's interest and personal life, and provides a means for users to socially interact over the internet, through e-mail, instant messaging or otherwise.

Recent years saw Facebook, Twitter, YouTube, LinkedIn, Skype, 9jabook, Lagbook and other social networking sites solidify their position at the heart of many users' daily internet activities, and saw these websites become a primary target for hackers and a vehicles for political revolutions in some countries (examples are Egypt, Tunisia, Libya and Saudi Arabia). Because of these, social networks have become one of the most significant vectors for data loss, identity theft and cyber terrorism (Sophos, 2010).

From an informational security standpoint, therefore, the pivotal weakness with social networking sites is conversely their strength: social networks encourage open interaction among both known users and loosely-connected users and, as a result, the normal social barriers against interacting with near strangers are lowered. Juxtapose this openness against the rampant increase in cyber crime and identity theft worldwide and therein lies a potential privacy and national security epidemic.

As of today, there are thousands of

*Privacy and National Security Issues in Social Networks: The Challenges*SNSs, with diverse technological affordances, supporting a wide range of interests and practices. While their key technological features are reasonably consistent, the cultures that emerge around SNSs varied. Most sites support the maintenance of pre-existing social networks, but others help visitors connect based on shared interests, political views, or activities. Some sites cater to diverse audiences, while others catch the attention of people based on common language or shared racial, political, sexual, religious, or nationality-based identities. Sites also vary in the extent to which they incorporate new information and communication tools, such as mobile connectivity, blogging, and photo/video-sharing.

## 2. Related Works

Participation in social networking sites has dramatically increased in recent years. Services such as Friendster, 9jabook, Lagbook, Hi5, Tribe, or the Facebook allow millions of individuals to create online profiles and share personal information with vast networks of friends - and, often, unknown numbers of strangers. Gross and Acquisti (2005) studied patterns of information revelation in online social networks and their privacy implications.

According to statistics published by some of the most well-known social networking services, there are more than 500 million active users on Facebook (Facebook, 2011), 175 million registered users on Twitter (Twitter, 2011), more than 100 million users on MySpace (MySpace, 2011), and more than 80 million members on LinkedIn (LinkedIn, 2011). Beyond Facebook and Twitter, social networking comprises a wide range of Web 2.0 tools. Public social networking media also include blogs, wikis, map-based mashups, and social news sites such as Digg.com and wikileaks.com (whistle blowers). Anyone can access these networks from work, home, or on the road. Users can disseminate any type of information, be it public or private, fact or fiction (PricewaterhouseCoopers, 2010).

Unsurprisingly, there have been countless reports of cyber criminals "phishing" for personal information on social networking sites (Networld, 2011). In fact, data suggest that an increasing volume of cyber crime is being directed to internet users on social networking sites. At risk is not only the personal information of the user, but presumably also that of the user's employer. The tools of the trade for cyber criminals are clever and devious, such as (i) creating fake profiles of friends, which is known as social engineering, (ii) hacking into friends' profiles and sending messages that look-and-feel to be from a friend, and (iii) emailing hostile computer code known as "malware," usually from an account of a "friend" that becomes activated when unwitting recipients click on the infected, internet links. Unsuspecting users on these sites run the risk of compromising sensitive information, including bank and financial data, highly personal information such as relationship, health and well-being and employment information, and similar sensitive information of family and/or friends (Esecurityplanet, 2011).

## 3. Types of Social Networks

SNSs can be divided in a number of ways. We have chosen to follow, with some extension, the division developed by Digizen (2008), an organisation which promotes secure activities on the web.

### 3.1 Profile-based Social Networks

Profile-based services are largely organised around members' profile pages. Bebo (www.bebo.com), Facebook (www.facebook.com), Hi5 (www.hi5.com), MySpace (www.myspace.com) and 9jabook.com are all good examples of this.

*International Journal of the Computer, the Internet and Management Vol. 19. No.3 (September-December, 2011) pp 14 -20*

15

### 3.2 Content-based Social Networks

With these services, the user's profile is the most important way of organising connections. Though, they play a secondary role in the posting of content. Photo-sharing site Flickr (www.flickr.com) is a good example of this brand of service, one where groups and comments are based around pictures. Shelfari (www.shelfari.com) is one of the current brand of book-focused sites, with the members 'bookshelf' being a focal point of their profile and membership.

### 3.3 White-label Social Networks

These sites present members with the opportunity to create and join communities – this means that users can build their own 'mini-MySpace's', small scale, personalised social networking sites about whatever the initiator wants them to be about. Some interesting examples are WetPaint (www.wetpaint.com), Wikileaks (www.wikileaks.com) and Wikipedia (www.wikipedia.com) which uses social wikis as its format to enable social networking.

### 3.4 Multi-User Virtual Environments

Gaming environments such as Runescape (www.runescape.com) and virtual world sites like Second Life (www.secondlife.com) allow users to interact with each other's avatars are a virtual representation of the user.

### 3.5 Mobile Social Networks

Many social networking sites are now offering mobile access to their services, allowing members to interact with their personal networks via their mobile phones. Two examples are Facebook (www.facebook.com) and Bebo (www.bebo.com). Increasingly, there are mobile-led and mobile-only based communities emerging, such as Glo AfriChart and MTN ToGo all in Nigeria.

### 3.6 Micro-blogging/Presence Updates

Many services let users post status updates i.e. short messages that can be updated to let people know what mood you are in or what you are doing. These types of networks enable users to be in constant touch with what their network is thinking, doing and talking about. Twitter (www.twitter.com), NairaLand (www.nairaland.com) and Wayn (www.wayn.com) are examples.

### 3.7 Social Search

Sites like Wink (www.wink.com), linkedIn (www.LinkedIn.com) and Spokeo (www.spokeo.com) produce results by searching across the public profiles of many social networking sites. This allows everyone to search by name, interest, location and other information available publicly on profiles, allowing the creation of web-based 'dossiers' on individuals.

### 3.8 Local Forums

Though often not included in social network definitions, place based fora such as Eastserve (www.eastserve.com), Onsnet (www.onsnetnuenen.nl), and Cybermoor (www.cybermoor.org) provide a localised form of social networking, linking online with offline activity.

### 3.9 Thematic Websites

Sites like Netmums (www.netmums.com) also include a local dimension by putting mums in touch with others in their locality, where they can share advice, information, recommendations, information on schools and are able to network both at the local and national levels. In addition, there are also sites for those with a disability such as www.deafgateway.info which provides a place for deaf people to interrelate.

## 4. Structure of Member Profiles

The member profile represents how the individual chooses to present their identity at a specific time and with a particular understanding of one's audience (Wildbit, 2005). But some observations show that while the audience and the individual evolve over time, one's profile is usually stuck in time. Some profile information are common to all social networking sites, these include Names (First Name, Surname or Middle Name) and e-mail contacts.

Table 1 shows the User profile information gathered by various social networking sites.

**Table 1**
*Structure of Profile Information in some Social Networks.*
*\* means the field exist*

| Profile Information | Facebook | Twitter | 9jabook | LinkedIn | Orkut | Friendster | Tribe | MySpace |
|---|---|---|---|---|---|---|---|---|
| Photo | * | * | * | * | * | * | * | * |
| Professional Detail | * | * | * | * | * |  | * |  |
| Gender | * | * | * | * | * | * | * | * |
| Age/Date of Birth | * | * | * | * | * | * | * | * |
| Sexual Orientation |  |  |  |  | * |  |  | * |
| Marital Status | * | * | * | * | * | * |  | * |
| Sense of Humor |  |  |  |  | * |  |  |  |
| Hobbies/Interest | * | * | * | * | * | * | * | * |
| Favorite Music | * |  | * | * | * | * | * | * |
| Favorite TV | * | * |  |  | * | * | * | * |
| Favorite Books | * | * |  |  | * | * | * | * |
| Favorite Food |  |  |  |  | * |  |  |  |
| Location | * | * | * | * | * | * | * | * |
| Home Town | * | * | * | * | * | * | * | * |
| Here for… |  |  |  | * | * | * | * | * |
| Schools | * | * | * | * |  | * | * | * |
| College/University | * | * | * | * |  | * |  |  |
| Clubs & Organizations | * | * | * |  |  |  | * |  |
| Languages | * | * | * | * |  |  | * |  |
| Religion | * | * | * | * | * |  |  | * |
| Smoking |  |  |  |  | * | * |  | * |
| Drinking |  |  |  |  | * | * |  | * |
| Nationality | * | * |  | * | * | * | * | * |

**Suggestions or Advice to Users:**
- The more profile information a user discloses or filled in his/her account, the more the user made public his/her private issues.
- Information which the user knows he/she uses as passwords or PIN (personal identification number) should remain blank. For example, some people use their date of birth or home town as passwords or PIN, in either case the field should be left black or made invisible.
- Users should also take full advantage of the feature that allows them to select which information in their profile will be visible to others.
- The closer potential connections between community members can be, the more information is the member willing to give about himself, beware!

## 5. The Privacy Issues

Attackers may use social networking services to spread malicious cryptogram, compromise users' computers, or access personal information concerning a user's identity, location, contact information, and personal or professional relationships. The user may also unintentionally reveal information to unofficial individuals by performing certain actions.

According to a Sophos survey conducted in December 2009, 60% of respondents believe that Facebook presents the biggest security risk of the social networking sites, significantly ahead of MySpace, Twitter and LinkedIn. The following are a few widespread threats to social networking services.

### 5.1 Viruses

The fame of social networking services makes them a perfect target for attackers who want to have the heaviest blow with the least effort. By creating a virus and embedding it in a website or a third-party application, an invader can potentially infect millions of computers just by relying on users to share the malicious links with their associates or friends.

### 5.2 Tools

Attackers may use tricks that allow them to take control of a user's account. The attacker could then have the right to use the user's private data and the data for any contacts that share their information with that user. An attacker with access to an account could also masquerade as that user and post malicious content.

### 5.3 Social Engineering Attacks

Attackers may send an email or post a statement that appears to originate from a trusted social networking service or user. The mail may contain a malicious URL or a demand for personal information. If you follow the instructions, you may reveal vital information or compromise the security of your system.

### 5.4 Identity Theft

Attackers may be able to collect enough personal information from social networking services to assume your identity or the identity of one of your friends. Even a few personal information may provide attackers with enough detail to guess answer to security or password reminder question for email, credit card, or bank accounts.

### 5.5 Third-party Applications

Some social networking services may let you to add third-party applications, including games and quizzes that provide additional functionality. Be cautious using these applications—even if an application does not contain malicious code, it might access information in your profile without your knowledge. This detail could then be used in a variety of ways, such as advertisements, performing market research, sending spam email, or accessing your contacts.

### 5.6 Business Data

Posting sensitive information intended only for internal company use on a social networking service can have serious consequences. Disclosing information about customers, intellectual property, human resource issues, mergers and acquisitions, or other company activities could result in liability or bad publicity, or could reveal information that is useful to competitors.

### 5.7 Professional Reputation

Inappropriate photos or content on a social networking service may threaten a user's educational and career prospects. Colleges and Universities may conduct online searches about potential students during the application process. Many companies also perform online searches of job candidates during the interview process. Information that suggests that a person might be unreliable, untrustworthy, or unprofessional could threaten the candidate's application. There have also been many instances of people losing their jobs for content posted to these services.

### 5.8 Personal Relationships

Because users can upload comments from any computer or smart phone that has internet access, they may impulsively post a comment that they later regret. According to a survey conducted by Retrevo, ―32 percent of people who post on a social networking site regret they shared information so openly. Even if comments and photos are retracted, it may be too late to undo the damage. Once information is online, there is no way to control who sees it, where it is redistributed, or what websites save it into their cache.

### 5.9 Personal Safety

The user may compromise personal security and safety by posting certain types of information on social networking services. For example, revealing that you will be away from home, especially if your address is posted in your profile, increases the risk that your home will be burglarized.

## 6. National Security Issues

Experts say cyber attack could come from anywhere—an individual or someone overseas, a terrorist group, or a country. But the number of ways a cyber attack could infiltrate national systems is growing—and the ever-expanding web of social networking sites could prove problematic for national cyber security. Social networking technologies are creating potential new challenges for government transparency and security, as more agency employees use Twitter, Facebook and similar external sites, officials at all levels of government are reviewing their policies.

Information posted to SNS sites can be a valuable resource for some, as targeted phishing attacks use validated information harvested from the web and identification checks used by legitimate sites. The danger of putting too much personal information online, particularly on social networking sites, was brought to light when the wife of the chief of the British secret service MI6 posted highly revealing details about their residence and friends on her Facebook page. This implies that when government officials became insecure, it means the nation is also insecure.

During the 2008 terrorist attacks in Mumbai, India for example, people on the scene sent Twitter updates, including the emergency contact number for the U.S. State Department's consular call center. The State Department's Deputy Assistant Secretary for Public Diplomacy used Twitter postings to provide updates on her personal experiences.

Recently in the Arab world countries who have spent years under the boot heel of a dictators have risen up and said "no more". Two decades back, in order to organize a resistance movement to an oppressive government one had to go outside and somehow find like-minded people (who were not secret police) and find a secluded place to quietly discuss matters. Even if you could get a significant revolt in a dictatorship one would most likely be killed or sent to a jell. The only way to get an oppressive regime to collapse back in the "old days" was for the economy to go belly up, the soldiers not getting paid, therefore would no longer defend the dictator (The Masked Walnut, 2011).

But thank God for the Internet social networking sites. One can now sit at home and talk to like minded people who are also sitting at the comfort of their homes. With Facebook and Twitter or even 9jabook, one can now organize thousands of people and mobilize them. Most of these people will not have met physically until they are actually at the protest. A tyrant can no longer "call out the tanks" because few minutes after they do their will be a live feed being broadcast on YouTube. If they are foolish enough to start shooting people their wouldn't be a head of a foreign government in their right mind not condemning the actions. If it can happen in Egypt and Tunisia it can happen anywhere, and all governments on earth know this. So, Nigerian government should be aware that online social networking constitutes great danger to national security!

## 7. Conclusion

Social networking is the current major trend in internet use. It is already heavily under attack and seems likely to continue to become more of a target as its popularity grows. Whether users eventually will be turned off by the rising tide of malware and spam may depend on how providers react and implement measures to ensure security and privacy.

For those of us that still wonder how the dictator of Tunisia was overthrown in less than one month after being in power for almost two and a half decades. There is no question about how opponents of his regime were able to topple it. Two words describe it: Facebook, Twitter. These two social networking sites enabled protesters to take to the streets, organize the opposition, recruit new protesters, and prevail over the police force and the military. What happens to National Security?

In conclusion, governments will have to play a major role in how secure social networks are, with much greater efforts required to crack down on current cybercriminals and discourage new blood from joining the dark side. These efforts must be implemented at both national and global level to ensure that crimes and criminals cannot be harbored and abetted by rogue nations ignoring global regulation. New laws must provide protection from criminals but also ensure secure behavior by those entrusted with sensitive data—who will doubtless continue to leak information in ever-greater amounts, as we have observed throughout the past decade.